\documentclass{agnspec}
\usepackage{graphics}
\usepackage{psfig}

\begin{document}
\title{On relativistic jet/counterjet asymmetry in the presence of radiating jet boundary layer }
\author{M. Ostrowski\inst{1} \and \L. Stawarz}
\institute{Obserwatorium Astronomiczne, Uniwersytet Jagiello\'nski,
 ul. Orla 171, 30-244 Krak\'ow, Poland\\
 \inst{1} mio@oa.uj.edu.pl }

\authorrunning{M. Ostrowski \& \L. Stawarz}
\titlerunning{On relativistic jet/counterjet asymmetry}
\maketitle

\begin{abstract}

We consider radiation of relativistic electrons accelerated within the jet extended boundary layer. Due to velocity shear across the boundary the observed jet/counterjet brightness ratio is diminished as compared to the one derived for the jet spine. Thus the jet Lorentz factor evaluated from the observed jet asymmetry can be an underestimated value influenced by observation of the slower boundary layer. We briefly discuss several consequences of the radiating boundary layer model in the context of recent Chandra and XMM observations of the large scale jets. 
%\vskip 0.2cm
\textbf{}
%\vskip 0.2cm

\end{abstract}

\section{Introduction}

Jet stratification was proposed in order to interpret radio and optical observations of large scale jets in radio galaxies (e.g., \cite{lai96}). The inferred jet morphology consists of a fast central {\it spine} surrounded by the {\it shear layer} extending into the broad jet {\it cocoon}. However, the physical properties of the boundary region are not exactly known. Polarimetry of the large scale jet boundaries usually show magnetic field being parallel to the jet axis, suggesting strong shearing effects at the jet edges. 3D numerical simulations reveal turbulent character of the boundary layer and its high specific internal energy (\cite{alo99}). Such regions are therefore promising places for particle acceleration, including both stochastic scattering in a turbulent medium and cosmic ray viscosity (cf. review by \cite{ber90}). With the parameters characteristic for the large scale jets, the former mechanism creates at the jet boundary a characteristic two-component electron spectrum: a power-law ended by a pile-up bump (\cite{ost00}, see also \cite{sta01}). A role of such electron distribution for the multiwavelength large scale jet emission was studied by \cite{sta02}.   

Recently, Chandra and XMM observatories detected significant X-ray nonthermal emission connected with the large scale jets in a number of radio loud AGNs. One of the most spectacular X-ray jets is the one observed in quasar PKS 0637-752 (\cite{cha00}). The most likely explanation of its X-ray emission is the inverse-Compton scattering of cosmic microwave background photons by the low energy nonthermal electrons (\cite{tav00}). This mechanism, however, requires highly relativistic flow velocities at the distance of order of 0.1 - 1 Mpc from the galactic nucleus. It is inconsistent with middly relativistic velocities inferred from the jet-counterjet radio brightness asymmetries of the large scale jets in radio galaxies. One may note, that the highly relativistic jets provide the most efficient way of energy transport to the terminal hot-spots in FR II sources (\cite{ghi01}). Therefore, there is striking disagreement between theoretical modeling and radio measurements for such sources. 

Here we discuss the possibility, that the emission of electrons accelerated within the turbulent shear layer affects the jet-counterjet brightness asymmetry observations, allowing the jet spine to remain highly relativistic even far away from the active nucleus. We point out several consequences of the presented model for radiative output of large scale jets, including their X-ray emission.

\section{Radiation of the jet boundary layer}

Let us consider (cf. \cite{sta02}) a large (tens-of-kpc) scale jet consisted of a highly relativistic cylindrical spine, with a radius $R_j \sim 1 \, {\rm kpc}$ and a bulk Lorentz factor $\Gamma_j$, surrounded by the turbulent boundary layer with a thickness $D \sim R_j$ and a radial velocity shear. We assume that the magnetic field with intensity $B \sim 10^{-5} \, {\rm G}$ is parallel {\it on average} to the flow velocity within the boundary region. The electrons injected into the shear layer are accelerated due to the resonant scattering on the magnetic field irregularities moving with, approximately, the Alfv\'en velocity, $V_A \sim 10^8 \, {\rm cm/s}$, and form a flat power-law energy spectrum, $\propto \gamma^{-\sigma}$. The acceleration time scale for this process is roughly $T_{acc} \sim r_g \, c \, V_A^{-2} \sim 10^2 \, \gamma \, {\rm [s]}$, where $r_g$ is the electron gyroradius, and $\gamma$ is its Lorentz factor. The electron escape from the shear layer with the assumed magnetic field configuration takes place due to cross-field diffusion, with the time scale much longer than the scale connected with synchrotron losses, $T_{loss} \sim 8 \cdot 10^{18} \, \gamma^{-1} \, {\rm [s]}$. As a result, the electrons form a pile-up bump at the maximum energy $\gamma_{eq} \sim 10^8$, where $T_{acc} \approx T_{loss}$. The slope of a power-law component and the normalization of the spectrum depend on physical conditions within the acceleration region. Below we assume energy equipartition between the magnetic field and relativistic electrons, providing electron spectrum normalization for $\sigma = 2$. For such condition, the radio-to-optical continuum is dominated by the emission of the power-law electron spectral component, $\nu^{-(\sigma-1)/2}$, while the high energy electron bump is pronounced at higher, X-ray frequencies. Thus the jet spectral properties at radio and optical frequencies can be different as compared to those at the X-rays, although they result from the synchrotron emission of the same electron population.

\begin{figure}
\centerline{\psfig{figure=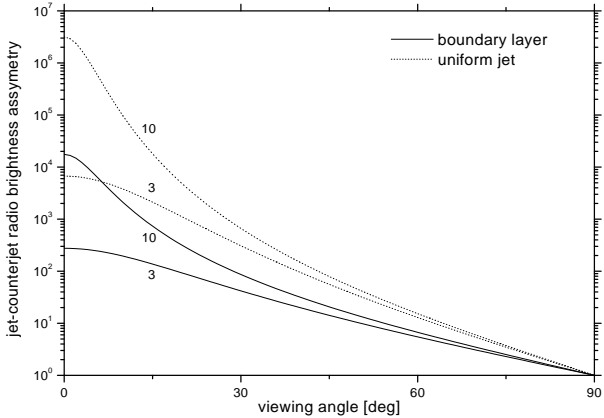,width=8.8cm,angle=0}}
\caption{ A ratio of the observed radio fluxes from the jet and the counterjet as a function of a viewing angle. The dotted lines correspond to the uniform jet models (`pure' spine radiation) with the bulk Lorentz factor $\Gamma_j = 10$ or $3$. The solid lines correspond to the radiation from a boundary shear layer with the assumed linear profile of the bulk Lorentz factor within the transition region, $1 < \Gamma < \Gamma_j$. The Lorentz factors $\Gamma_j$ are provided near the respective curves. }
\end{figure}

An important manifestation of the boundary layer emission is a decrease of the jet-counterjet brightness asymmetry as compared to the uniform jet models, suggested previously by \cite{kom90}. In the presence of the radiating shear layer, an observer with a line of sight inclined at angle $\theta > 1 \, / \, \Gamma_j$ with respect to the jet axis will see predominantly radiation generated within the transition region, $1 < \Gamma < \Gamma_j$, while the spine emission will be Doppler-hidden. Note, that the spine and the boundary layer radiation can differ not only due to the kinematic effects involved, but also because of a possible presence of distinct electron populations within those two regions. Figure 1 illustrates for different values of $\Gamma_j$ the boundary layer jet-counterjet brightness asymmetry for the spectral index $ (\sigma -1 )/2 = 0.5$ and a linear Lorentz factor profile within the boundary region. For comparison, the uniform jet models are also presented. Due to the considered effect, for $\Gamma_j = 10$ and the moderate inclinations, the observed asymmetry can be reduced by more than one order of magnitude in comparison to the uniform jets. 

\section{Discussion}

Relativistic velocities of large scale jets (corresponding to the bulk Lorentz factor $\Gamma_j \sim 10$) were postulated in order to explain X-ray emission detected by Chandra and XMM from the jets with small inclination angles. Much smaller velocities of the jets in radio galaxies inferred from the jet-counterjet radio brightness asymmetry can be explained as a result of the radiating boundary layer with the velocity shear. The jet spines flowing with large bulk Lorentz factors can carry high kinetic energy, $L_{tot} \sim 10^{47} {\rm erg/s}$, assuming that they are composed from `normal' electron-proton plasma (cf. \cite{ghi01}), and efficiently produce X-rays due to Compton scattering of CMB photons. The radiating boundary layer can dominate the X-ray emission of the jets observed at large viewing angles, as discussed by \cite{cel01} and \cite{sta02}. In the later model, such X-ray emission is produced by synchrotron radiation of the high energy electron pile-up bump. Therefore, the jet-counterjet X-ray brightness asymmetry differ from the radio one due to different spectral characters of the power-law and the spectral bump emission. One should note, that recent HST observations of the jet in 3C 273 (\cite{jes02}) show spectral flattening starting at UV frequencies and suggest that our interpretation of the large scale jets' X-ray emission can be correct at least for some sources.   
 
\begin{acknowledgements}
The present work was supported by Komitet Bada\'{n} Naukowych through the grant BP 258/P03/99/17. 
\end{acknowledgements}

\end{document}